\documentclass[twocolumn,english,american]{revtex4-2}
\usepackage[T1]{fontenc}
\usepackage[utf8]{luainputenc}
\usepackage{amsmath}
\usepackage{amssymb}
\usepackage{babel}
\begin{document}
\title{Projected Augmented Waves (PAW): extended resolution of unity method}
\author{Garry Goldstein}
\address{garrygoldsteinwinnipeg@gmail.com}
\begin{abstract}
The Projected Augmented Waves (PAW) method is based on a linear transformation
between the pseudo wavefunctions and the all electron wavefunctions.
To obtain high accuracy with this method, it is important that the
local part of the linear transform (inside each atomic sphere) be
defined over a complete basis set (with deviations from completeness
leading to corrections to the total energy not computed within current
implementations of PAW). Here we show how to make this basis much
closer to complete without significant additional computational work
and without modifying the transformation in any significant way thereby
making the modifications we propose easy to implement in current electronic
structure codes for PAW.. This is done by extending the resolution
of unity used for the transform to include more smooth wavefunctions
(which have nothing to do with the atomic problem) and having them
linearly transform via the identity.
\end{abstract}
\maketitle

\section{Introduction}\label{sec:Introduction}

It is well know that all electron methods are often more technically
difficult to practically implement then pseudopotential methods \citep{Wills_2010,Soler_1990,Soler_1989,Slater_1937,Sjostedt_2000,Singh_2006,Singh_1991,Michalicek_2014,Michalicek_2013,Martin_2020,Loucks_1967,Andersen_2003,Andersen_1984,Andersen_1975}.
This is in part due to the complex nature of the basis wavefunctions,
for all electron methods, which are often based on augmentation methods
- where the wavefunction has markedly different expressions in the
Muffin Tin (MT) regions and in the interstitial regions respectively.
This augmentation construction often complicates the evaluation of
the Khon Sham (KS) Hamiltonian and related quantities needed to solve
electronic structure problems. This leads to some technical disadvantages
of all electron methods vs. pseudopotential ones, in particular: 
\begin{itemize}
\item For all electron methods the calculation of eigenvalues and eigenvectors
of the Khon Sham (KS) Hamiltonian often requires FLAPW (Full Potential
Linearized Augmented Plane Waves \citep{Blaha_1990,Hamann_1979,Jansen_1984,Mattheis_1986,Skriver_1984,Wei_1985,Wei_1985(2),Wimmer_1981})
which requires a relatively complex setup with multiple steps and
calculations. On the other hand full potential plane wave methods
have simple expressions for KS Hamiltonian matrix elements and reduce
to the identity matrix for wavefunction overlaps in many cases - in
particular for norm conserving pseudopotentials.
\item The Coulomb term (Hartree energy) is more complex in augmented, all
electron, systems. Indeed it requires the Weinert method for efficient
calculation and convergence without shape approximations \citep{Weinert_1981,Blugel_2006}.
Where as for pseudopotentials Ewald sums suffice for most applications
\citep{Marx_2009}. 
\item Atomic force calculations in all electron methods are more complex
as they require the computation of the Pulay contribution to the forces
- which arise due to changes of span of the all electron basis set
with atomic positions (the Hellman-Feynman theorem does not apply
directly in this case \citep{Yu_1991} and needs modifications), while
many pseudized methods require only the Hellman-Feynman terms.
\item Substitutional alloys are difficult to study for all electron methods
(other then through cumbersome methods such as Korringa Khon Rostoker
(KKR) methods \citep{Korringa_1947,Khon_1954} or through supercells),
where as for pseudopotentials there are many viable virtual crystal
approximation methods \citep{Bellaiche_2000,Steiner_2016}.
\item The Car Parrinello molecular dynamics simulations \citep{Car_1985,Marx_2009,Blochl_1992,Laasonen_1993}
method is significantly easier with pseudopotentials then with augmented
waves such as Linearized Augmented Plane Waves (LAPW) and very limited
work has been done so far with all electron methods in that direction
\citep{Marx_2009,Singh_2006}.
\end{itemize}
As such, despite many advances in the accuracy and efficiency of all
electron methods and their conceptual simplicity \citep{Goldstein_2024(3),Goldstein_2024(2),Goldstein_2024,Singh_1991,Sjostedt_2000,Michalicek_2013,Michalicek_2014,Soler_1989,Soler_1990,Goldstein_2024(4)},
it is extremely worthwhile to study pseudopotential methods (atleast
currently) - as they are currently the leading methods for electronic
structure calculations within Density Functional Theory (DFT). Most
modern pseudopotential methods may be loosely divided into three different
categories:
\begin{itemize}
\item Norm conserving pseudopotentials \citep{Martin_2020,Marx_2009}
\item Ultrasoft psudopotentials \citep{Vanderbilt_1990,Martin_2020}
\item Projected Augmented Waves (PAW) \citep{Blochl_1994,Rostgaard_2009,Kresse_1999}
\end{itemize}
We now describe each of these three electronic structure methods in
turn. 

Norm conserving pseudopotentials are based on the idea that a perfectly
spherical scattering potential (the atomic sphere, when small enough,
is roughly spherically symmetric) can be represented by a scattering
phase shift - in each partial wave and at each energy - as a boundary
condition at the surfaces of the sphere. Any potential producing these
scattering phase shifts is sufficient for calculations of eigenvalues
of the KS Hamiltonian - not just the original atomic one. Therefore
an equivalent problem where a much smoother potential with the same
scattering phase shift (in the middle of the valence band) as the
atomic potential, in each different angular momentum channel, can
be used in the solution of the KS Hamiltonian. These smoother potentials
may be used to solve for the eigenvalues of the KS problem leading
to lower cutoffs (smoother wavefunctions) needed for electronic structure
calculations. The norm conserving condition insures both accurate
Madelung energy and that the derivative of the scattering phase shift
with respect to the energy is the same both for the atomic potential
and the pseudopotential leading to improved accuracy as a function
of energy (better transferability). 

A competitor to norm conserving pseudopotentials are ultrasoft potentials.
These pseudopotentials use the many state generalizations of the Kleinman-Bylander
transform \citep{Martin_2020} in order to reproduce the scattering
phase shifts of the atomic problem simultaneously at many energies.
This allows one to get rid of the norm conserving condition (needed
for accuracy and transferability without multiple scattering energies)
and restore correct Madelung energies and with a fictitious overlap
matrix between the pseudized wavefunctions and compensating charges
(for higher multipoles). This greatly softens the potential as the
pseudopotential need not conserve norm and therefore has fewer constraints
leading to softer pseudopotentials, e.g. especially for $3d$ transition
metals which are hard to soften while conserving norm as the $3d$
wavefunctions have no oscillations to smooth out with a pseudized
wavefunction \citep{Marx_2009,Martin_2020,Vanderbilt_1990}. This
also solves the problem of ghost bands: semi-core bands with significantly
different scattering phase shifts then in the middle of the valence
band that often plagued norm conserving pseudopotentials can now be
handled by introducing a Kleinman-Bylander transform with a state
at the energy of the semi-core states thereby insuring good wavefunctions
there. The price for this is minor and associated with some additional
terms in the KS equations which come about because of the fictitious
overlap (norm). 

For PAW we introduce a linear transformation between the soft (low
cutoff) pseudized wavefunctions and the all electron wavefunctions.
The various quantities needed to set up the KS equation and obtain
the ground state energy are then expressed through this linear transformation.
PAW is sometimes considered the most versatile because at its core
its an efficient method to map a pseudo-wave function problem onto
an all electron problem - leading to great accuracy of computations.
One of its limitations is that it assumes some completeness relations
for pseudo wavefunctions used in the transform, a problem we here
partially remedy at very low computational cost and in a manner that
is highly compatible with current electronic structure codes for PAW. 

\section{PAW Summary}\label{sec:PAW-Summary}

We now describe PAW in detail. First we note that the all electron
wavefunction is given through the relationship: 
\begin{equation}
\psi=\mathcal{T}\tilde{\psi}\label{eq:Transform-1}
\end{equation}
where $\tilde{\psi}$ is the a pseudo wavefunction. We now write:
\begin{equation}
\mathcal{T}=\mathbb{I}+\sum_{\mathbf{R}}S_{\mathbf{R}}\label{eq:Decomposition}
\end{equation}
Here $\mathbf{R}$ are the sites of the atomic nuclei and $S_{\mathbf{R}}$
is a linear transformation localized at the augmentation sphere centered
at $\mathbf{R}$. More precisely:
\begin{equation}
S_{\mathbf{R}}=\sum_{i\in\mathbf{R}}\left[\left|\phi_{i}\right\rangle -\left|\tilde{\phi}_{i}\right\rangle \right]\left\langle \tilde{p}_{i}\right|\label{eq:Definition}
\end{equation}
Where $\left|\phi_{i}\right\rangle $ are from atomic calculations
representing the exact atomic states while $\left|\tilde{\phi}_{i}\right\rangle $
are smooth functions localized inside the augmentation spheres. Where:
\begin{equation}
\left\langle \tilde{p}_{i}\mid\tilde{\phi}_{j}\right\rangle =\delta_{ij}\label{eq:Identity-2}
\end{equation}
There are many options for the wavefunctions $\left|\tilde{p}_{i}\right\rangle $,
for example: 
\begin{equation}
\left|\tilde{p}_{i}\right\rangle =\sum_{j\in\mathbf{R}}\left|\tilde{\phi}_{j}\right\rangle \left[M\right]_{ji}^{-1}\label{eq:Projectors}
\end{equation}
In this work we will consider only this case (though similar considerations
may be made for more general cases \citep{Rostgaard_2009}). Where
\begin{equation}
M_{ij}=\left\langle \tilde{\phi}_{i}\mid\tilde{\phi}_{j}\right\rangle \label{eq:Inverse}
\end{equation}
This means that 
\begin{equation}
\left[\sum_{i\in\mathbf{R}}\left|\tilde{\phi}_{i}\right\rangle \left\langle \tilde{p}_{i}\right|\right]\left|\tilde{\phi}_{j}\right\rangle =\left|\tilde{\phi}_{j}\right\rangle \label{eq:Identity}
\end{equation}
so that 
\begin{equation}
\sum_{i\in\mathbf{R}}\left|\tilde{\phi}_{i}\right\rangle \left\langle \tilde{p}_{i}\right|=\mathbb{I}_{\mathbf{R}}\label{eq:Identity-1}
\end{equation}
provided that the $\left|\tilde{\phi}_{i}\right\rangle $ form a complete
basis set. However we have that in reality:
\begin{equation}
\sum_{i\in\mathbf{R}}\left|\tilde{\phi}_{i}\right\rangle \left\langle \tilde{p}_{i}\right|=\left\{ \begin{array}{cc}
\mathbb{I}_{\mathbf{R}} & \in Span\left\{ \left|\tilde{\phi}_{i}\right\rangle \right\} \\
0 & otherwise
\end{array}\right.\label{eq:Span}
\end{equation}
This relation, assumed completeness Eq. (\ref{eq:Identity-1}), is
used to prove the identity that for a local operator: 
\begin{equation}
\left\langle \tilde{\psi}\right|\hat{A}\left|\tilde{\psi}\right\rangle _{\mathbf{R}}=\left\langle \tilde{\psi}_{\mathbf{R}}\right|\hat{A}\left|\tilde{\psi}_{\mathbf{R}}\right\rangle \label{eq:Identity-3}
\end{equation}
Where $\left\langle \right\rangle _{\mathbf{R}}$ means all integrations
are done over the atomic sphere centered about $\mathbf{R}$. Where:
\begin{equation}
\left|\psi_{\mathbf{R}}\right\rangle =\left[\sum_{i\in\mathbf{R}}\left|\phi_{i}\right\rangle \left\langle \tilde{p}_{i}\right|\right]\left|\tilde{\psi}\right\rangle \label{eq:Definition-1}
\end{equation}
These relations are used to compute the form of the density of electrons
the correlation and exchange and the Hartree (Coulomb piece) of the
Hamiltonian \citep{Blochl_1994,Martin_2020,Marx_2009}. This is essential
to current PAW methodology \citep{Rostgaard_2009}. Here we improve
on this resolution of identity without changing its mathematical form
or incurring significant additional computational costs.

\section{Main Construction}\label{sec:Main-Construction}

We now address the main problem pointed out in the previous section
that the $\left|\tilde{\phi}_{i}\right\rangle $ do not form a complete
basis set. This is in particular because the set $\left|\tilde{\phi}_{i}\right\rangle $
of the same cardinality as $\left|\phi_{i}\right\rangle $ and we
must obtain $\left|\phi_{i}\right\rangle $ from atomic calculations,
where typically only two or three wavefunctions, per angular momentum
channel, are available with current PAW atomic data sets \citep{Rostgaard_2009}.
Here we propose to circumvent this bottleneck by introducing: 
\begin{equation}
\psi=\mathcal{T}_{M}\tilde{\psi}\label{eq:Transform_Multy}
\end{equation}
where $\tilde{\psi}$ is the a pseudo wavefunction and $M$ stands
for multiple states. We now write: 
\begin{equation}
\mathcal{T}_{M}=\mathbb{I}+\sum_{\mathbf{R}}S_{\mathbf{R}}^{M}\label{eq:Definition_multy}
\end{equation}
Where 
\begin{equation}
S_{\mathbf{R}}^{M}=\sum_{i,\alpha\in\mathbf{R}}\left[\left|\phi_{i\alpha}\right\rangle -\left|\tilde{\phi}_{i\alpha}\right\rangle \right]\left\langle \tilde{p}_{i\alpha}\right|\label{eq:Main_idea}
\end{equation}
Where 
\begin{equation}
\left|\phi_{i\alpha}\right\rangle =\left\{ \begin{array}{cc}
\left|\phi_{i}\right\rangle  & \alpha=1\\
\left|\tilde{\phi}_{i\alpha}\right\rangle  & \alpha>1
\end{array}\right.\label{eq:Definitions}
\end{equation}
Here $\alpha$ runs over a list of on the order of ten to one hundred
additional pseudo wavefunctions per an all electron wavefunction.
Where we have used the idea that it is much easier to generate smooth
(low cutoff) wave functions - say by a combination of Bessel functions
with different energies multiplied by spherical harmonics with different
angular momentum channels, then to do atomic data set calculations.
These wavefunctions can be chosen to have good overlaps with the plane
waves (Bessel functions) used for the basis for the KS problem for
PAW in various angular momentum channels and therefore can be chosen
as a nearly complete basis set for the space of the pseudized wave
functions considered in the PAW method. As such multiple smooth wavefunctions
can easily be added to the same all electron atomic calculation wavefunction.
Where we have that:
\begin{equation}
\left\langle \tilde{p}_{i\alpha}\mid\tilde{\phi}_{j\beta}\right\rangle =\delta_{ij}\delta_{\alpha\beta}\label{eq:Identity-4}
\end{equation}
There are many options for the wavefunctions $\left|\tilde{p}_{i\alpha}\right\rangle $
for example: 
\begin{equation}
\left|\tilde{p}_{i\alpha}\right\rangle =\sum_{j\in\mathbf{R}}\left|\tilde{\phi}_{j\beta}\right\rangle \left[M\right]_{j\beta i\alpha}^{-1}\label{eq:Projectors-1}
\end{equation}
Where 
\begin{equation}
M_{i\alpha j\beta}=\left\langle \tilde{\phi}_{i\alpha}\mid\tilde{\phi}_{j\beta}\right\rangle \label{eq:Inverse-1}
\end{equation}
We notice that: 
\begin{equation}
\sum_{i\in\mathbf{R}}\left|\tilde{\phi}_{i\alpha}\right\rangle \left\langle \tilde{p}_{i\alpha}\right|=\left\{ \begin{array}{cc}
\mathbb{I}_{\mathbf{R}} & \in Span\left\{ \left|\tilde{\phi}_{i\alpha}\right\rangle \right\} \\
0 & otherwise
\end{array}\right.\label{eq:Span-1}
\end{equation}
So is much closer to the resolution of identity, as such we have inserted
the identity transform in a clever way improving the correctness of
the resolution of identity.

\section{Some technical details}\label{sec:Some-technical-details}

In this section we present the main technical consequences of our
new transform as the apply to current implementations of PAW. However
we note that since the mathematical structUre of our resolution of
identity is identical to the one usually used for PAW these technical
consequences would already be coded in electronic structure codes
for PAW and require no further work to implement.

\subsection{Main modifications to local operators}\label{sec:Main-idea-1}

We now write:
\begin{align}
\left|\psi_{\mathbf{R}}^{M}\right\rangle  & =\left[\sum_{i\alpha\in\mathbf{R}}\left|\phi_{i\alpha}\right\rangle \left\langle \tilde{p}_{i\alpha}\right|\right]\left|\tilde{\psi}\right\rangle \nonumber \\
\left|\tilde{\psi}_{\mathbf{R}}^{M}\right\rangle  & =\left[\sum_{i\alpha\in\mathbf{R}}\left|\tilde{\phi}_{i\alpha}\right\rangle \left\langle \tilde{p}_{i\alpha}\right|\right]\left|\tilde{\psi}\right\rangle \label{eq:Wavefunctions-1}
\end{align}
Now for a local operator $\hat{A}$ we write: 
\begin{align}
 & \left\langle \mathcal{T}_{M}^{\dagger}\hat{A}\mathcal{T}_{M}\right\rangle \nonumber \\
 & =\left[\left\langle \tilde{\psi}\right|+\sum_{\mathbf{R}}\left[\left\langle \psi_{\mathbf{R}}^{M}\right|-\left\langle \tilde{\psi}_{\mathbf{R}}^{M}\right|\right]\right]\times\nonumber \\
 & \times\hat{A}\left[\left|\tilde{\psi}\right\rangle +\sum_{\mathbf{R}}\left[\left|\psi_{\mathbf{R}}^{M}\right\rangle -\left|\tilde{\psi}_{\mathbf{R}}^{M}\right\rangle \right]\right]\nonumber \\
 & =\left[\left\langle \tilde{\psi}\right|-\sum_{\mathbf{R}}\left\langle \tilde{\psi}_{\mathbf{R}}^{M}\right|\right]\hat{A}\left[\left|\tilde{\psi}\right\rangle -\sum_{\mathbf{R}'}\left|\tilde{\psi}_{\mathbf{R}'}^{M}\right\rangle \right]+\nonumber \\
 & +\sum_{\mathbf{R}}\left\langle \psi_{\mathbf{R}}^{M}\right|\hat{A}\left|\psi_{\mathbf{R}}^{M}\right\rangle \nonumber \\
 & =\left\langle \tilde{\psi}\right|\hat{A}\left|\tilde{\psi}\right\rangle +\sum_{\mathbf{R}}\left[\left\langle \psi_{\mathbf{R}}^{M}\right|\hat{A}\left|\psi_{\mathbf{R}}^{M}\right\rangle -\left\langle \tilde{\psi}_{\mathbf{R}}^{M}\right|\hat{A}\left|\tilde{\psi}_{\mathbf{R}}^{M}\right\rangle \right]\label{eq:Expecation}
\end{align}
It is now also useful to introduce:
\begin{equation}
D_{i\alpha j\beta}^{\mathbf{R}}=\sum_{n,\mathbf{k}}f\left(\varepsilon_{n}\left(\mathbf{k}\right)\right)\left\langle \tilde{\psi}_{n}\left(\mathbf{k}\right)\mid\tilde{p}_{i\alpha}\right\rangle \left\langle \tilde{p}_{j\beta}\mid\tilde{\psi}_{n}\left(\mathbf{k}\right)\right\rangle \label{eq:Weights}
\end{equation}
This means that: 
\begin{align}
 & \left\langle \mathcal{T}_{M}^{\dagger}\hat{A}\mathcal{T}_{M}\right\rangle \nonumber \\
 & =\sum_{n,\mathbf{k}}f\left(\varepsilon_{n}\left(\mathbf{k}\right)\right)\left\langle \tilde{\psi}_{n}\left(\mathbf{k}\right)\right|\hat{A}\left|\tilde{\psi}_{n}\left(\mathbf{k}\right)\right\rangle +\nonumber \\
 & +\sum_{\mathbf{R},i,j}D_{i\alpha j\beta}^{\mathbf{R}}\left[\left\langle \phi_{i\alpha}\right|\hat{A}\left|\phi_{j\beta}\right\rangle -\left\langle \tilde{\phi}_{i\alpha}\right|\hat{A}\left|\tilde{\phi}_{j\beta}\right\rangle \right]\nonumber \\
 & +\sum_{core\in\mathbf{R}}\left\langle \phi_{c,i}\right|\hat{A}\left|\phi_{c,i}\right\rangle \label{eq:Single_particle}
\end{align}
Here we have introduced core states from the atomic calculations $\left|\phi_{c,i}\right\rangle $.
Notice that this is a different expression then the one with regular
PAW in particular there are cross terms now between $\alpha=1$ and
$\alpha>1$. In particular works for 
\begin{equation}
\hat{A}=\delta\left(\mathbf{r}-\mathbf{r}'\right),\:\hat{A}=\frac{-\nabla^{2}}{2m}\label{eq:Density}
\end{equation}
which are important for ground state energy determination. That is
for the density and kinetic energy terms. 

\subsection{Main modification to KS equations}\label{subsec:Main-modification-to}

We note that: 
\begin{equation}
\mathcal{T}_{M}^{\dagger}\hat{H}_{KS}\mathcal{T}_{M}\left|\tilde{\psi}_{n}\right\rangle =\varepsilon_{n}\mathcal{T}_{M}^{\dagger}\mathcal{T}_{M}\left|\tilde{\psi}_{n}\right\rangle \label{eq:KS_equation-1}
\end{equation}
Furthermore: 
\begin{align}
 & S_{M}\equiv\mathcal{T}_{M}^{\dagger}\mathcal{T}_{M}=1+\sum_{\mathbf{R}}S_{\mathbf{R}}^{M}+\sum_{\mathbf{R}}S_{\mathbf{R}}^{M\dagger}+\sum_{\mathbf{R}}S_{\mathbf{R}}^{M\dagger}S_{\mathbf{R}}^{M}\nonumber \\
 & =1+\left[\sum_{i\alpha\in\mathbf{R}}\left|\tilde{p}_{i\alpha}\right\rangle \left\langle \tilde{\phi}_{i\alpha}\right|\right]\sum_{i,\alpha\in\mathbf{R}}\left[\left|\phi_{j\beta}\right\rangle -\left|\tilde{\phi}_{j\beta}\right\rangle \right]\left\langle \tilde{p}_{j\beta}\right|+\nonumber \\
 & +\sum_{i,\alpha\in\mathbf{R}}\left|\tilde{p}_{i\alpha}\right\rangle \left[\left\langle \phi_{i\alpha}\right|-\left\langle \tilde{\phi}_{i\alpha}\right|\right]\left[\sum_{j\beta\in\mathbf{R}}\left|\tilde{\phi}_{j\beta}\right\rangle \left\langle \tilde{p}_{j\beta}\right|\right]+\nonumber \\
 & +\sum_{i,\alpha\in\mathbf{R}}\left|\tilde{p}_{i\alpha}\right\rangle \left[\left\langle \phi_{i\alpha}\right|-\left\langle \tilde{\phi}_{i\alpha}\right|\right]\sum_{j\beta\in\mathbf{R}}\left[\left|\phi_{j\beta}\right\rangle -\left|\tilde{\phi}_{j\beta}\right\rangle \right]\left\langle \tilde{p}_{j\beta}\right|\nonumber \\
 & =1+\sum_{i,\alpha\in\mathbf{R}}\sum_{j\beta\in\mathbf{R}}\left|\tilde{p}_{i\alpha}\right\rangle \left[\left\langle \phi_{i\alpha}\mid\phi_{j\beta}\right\rangle -\left\langle \tilde{\phi}_{i\alpha}\mid\tilde{\phi}_{j\beta}\right\rangle \right]\left\langle \tilde{p}_{j\beta}\right|\nonumber \\
 & \equiv1+\sum_{i,\alpha\in\mathbf{R}}\sum_{j\beta\in\mathbf{R}}\left|\tilde{p}_{i\alpha}\right\rangle \Delta_{i\alpha j\beta}\left\langle \tilde{p}_{j\beta}\right|\neq\mathcal{T}^{\dagger}\mathcal{T}=S\label{eq:S_overlap}
\end{align}
The reason for the inequality is that we have used a different (more
accurate) resolution of identity to simplify the situation which leads
to different results, in particular note again the cross terms between
$\alpha=1$ and $\alpha>1$. Similarly for:
\begin{align}
 & \mathcal{T}_{M}^{\dagger}\hat{H}_{KS}\mathcal{T}_{M}=\nonumber \\
 & =\sum_{i,\alpha\in\mathbf{R}}\sum_{j\beta\in\mathbf{R}}\left[\left\langle \phi_{i\alpha}\mid\hat{H}_{KS}\mid\phi_{j\beta}\right\rangle -\left\langle \tilde{\phi}_{i\alpha}\mid\hat{H}_{KS}\mid\tilde{\phi}_{j\beta}\right\rangle \right]\times\nonumber \\
 & \times\left|\tilde{p}_{i\alpha}\right\rangle \left\langle \tilde{p}_{j\beta}\right|+\hat{H}_{KS}\nonumber \\
 & \equiv\hat{H}_{KS}+\sum_{i,\alpha\in\mathbf{R}}\sum_{j\beta\in\mathbf{R}}\left|\tilde{p}_{i\alpha}\right\rangle h_{i\alpha j\beta}\left\langle \tilde{p}_{j\beta}\right|\neq\mathcal{T}^{\dagger}\hat{H}_{KS}\mathcal{T}\label{eq:KS_Hamiltonian}
\end{align}
Again the resolution of identity changed the final result. Here $\hat{H}_{KS}=\nabla^{2}+V_{KS}$
where $V_{KS}$ is the KS potential. Furthermore the resolution of
identity is used repeatedly for finding the ground state energy (see
Section \ref{subsec:Correlation-and-exchange}) and therefore also
its derivative with respect to the electron density which is the value
of the KS potential.

\subsection{Correlation and exchange and Hartree piece}\label{subsec:Correlation-and-exchange}

We write for local correlation and exchange functionals such as Local
Density Approximation (LDA) or Generalized Gradient Approximation
(GGA):
\begin{equation}
E_{XC}=E_{XC}\left(\tilde{n}\right)+\sum_{\mathbf{R}}E_{XC}\left(n_{\mathbf{R}}\right)-\sum_{\mathbf{R}}E_{XC}\left(\tilde{n}_{\mathbf{R}}\right)\label{eq:Correlation}
\end{equation}
This is based on the idea that 
\begin{equation}
\tilde{n}=\tilde{n}_{\mathbf{R}}\:\left(\mathrm{in\:the\:sphere}\:\mathbf{R}\right)\label{eq:correlation}
\end{equation}
However for this we see that Eq. (\ref{eq:Span-1}) is much better
then Eq. (\ref{eq:Span}) with explicit changes in the form of $\tilde{n}_{\mathbf{R}}$
seen from Eq. (\ref{eq:Single_particle}). Furthermore we now introduce
the convenient notation:
\begin{equation}
\left(f\mid g\right)=\int d^{3}\mathbf{r}_{1}\int d^{3}\mathbf{r}_{2}\frac{f\left(\mathbf{r}_{1}\right)g\left(\mathbf{r}_{2}\right)}{\left|\mathbf{r}_{1}-\mathbf{r}_{2}\right|}\label{eq:Convenient}
\end{equation}
and 
\begin{equation}
\left(f\mid f\right)=\left(\left(f\right)\right)\label{eq:short_hand}
\end{equation}
Similarly we have that the Coulomb piece is given by:
\begin{align}
E_{C}\left[n\right] & =U_{H}\left(\tilde{\rho}\right)\nonumber \\
 & +\frac{1}{2}\sum_{\mathbf{\mathbf{R}}}\left[\left(\left(n_{\mathbf{R}}\right)\right)+2\left(n_{\mathbf{R}}\mid Z_{\mathbf{R}}\right)-\left(\left(\tilde{n}_{\mathbf{R}}+\tilde{Z}_{\mathbf{R}}\right)\right)\right]\label{eq:Coulomb-1}
\end{align}
where 
\begin{equation}
U_{H}\left(\tilde{\rho}\right)=\frac{1}{2}\left(\left(\tilde{n}+\tilde{Z}_{\mathbf{R}}\right)\right)\label{eq:Hartree}
\end{equation}
Where $Z_{\mathbf{R}}$ is the coulomb charge density and $\tilde{Z}_{\mathbf{R}}$
is the pseudized charge density satisfying the relationship: 
\begin{align}
 & \int d^{3}\mathbf{r}\left|\mathbf{r}-\mathbf{R}\right|^{l}Y_{lm}\left(\widehat{\mathbf{r}-\mathbf{R}}\right)\times\nonumber \\
 & \times\left(n_{\mathbf{R}}\left(\mathbf{r}\right)-\tilde{n}_{\mathbf{R}}\left(\mathbf{r}\right)+Z_{\mathbf{R}}\left(\mathbf{r}\right)-\tilde{Z}_{\mathbf{R}}\left(\mathbf{r}\right)\right)\nonumber \\
 & =0\label{eq:Vansih}
\end{align}
Where we have repeatedly used Eq. (\ref{eq:correlation}) which is
dependent on the resolution of unity (see Section \ref{sec:Main-idea-1})
so Eq. (\ref{eq:Span-1}) is much better then Eq. (\ref{eq:Span})
for total energy calculations.

\section{Conclusion}\label{sec:Conclusion}

In this work we have introduced a new version (extended basis set)
of the PAW transformation (given in Eqs. (\ref{eq:Transform_Multy}),
(\ref{eq:Definition_multy}), (\ref{eq:Main_idea}) and (\ref{eq:Definitions}))
which allows for an extended resolution of identity as part of the
map between the pseudized problem and the all electron problem. This
greatly improves the accuracy with which mapping reproduces the key
PAW equation for expectation values local operators (given in Eq.
(\ref{eq:Identity-3})) as well as the various applications of that
identity to exchange and correlation energies, Coulomb energies and
the KS secular equation. This improves the accuracy of PAW at very
limited computational cost and because of the mathematical form of
our equations is nearly identical to that of regular PAW it can readily
be incorporated into modern electronic structure implementations of
PAW. In the future it would be of interest to include strong correlations
into the calculations in the form of LDA+U or LDA+DMFT or LDA+GA \citep{Kotliar_2006,Lee_2024}
with this new resolution of identity for the PAW method.
\selectlanguage{english}%

\end{document}